\documentclass[draft]{agujournal2019}
\usepackage{url}
\usepackage{lineno}
\usepackage[inline]{trackchanges}
\usepackage{soul}
\draftfalse
\journalname{Geophysical Research Letters}

\begin{document}

\title{The dimmest state of the Sun}
\authors{K. L. Yeo\affil{1}, S. K. Solanki\affil{1,2}, N. A. Krivova\affil{1}, M. Rempel\affil{3}, L. S. Anusha\affil{1}, A. I. Shapiro\affil{1}, R. V. Tagirov\affil{4,1}, V. Witzke\affil{1}}
\affiliation{1}{Max Planck Institute for Solar System Research, 37077 G\"ottingen, Germany}
\affiliation{2}{School of Space Research, Kyung Hee University, Yongin, Gyeonggi 446-701, Republic of Korea}
\affiliation{3}{High Altitude Observatory, National Center for Atmospheric Research, Boulder, CO 80307, USA}
\affiliation{4}{Blackett Laboratory, Imperial College London, London SW7 2AZ, UK}
\correspondingauthor{K. L. Yeo}{yeo@mps.mpg.de}

\begin{keypoints}
\item The role of solar forcing in global warning is unclear due to the uncertainty in the change in TSI since the Maunder minimum 
\item The TSI level of the Sun when it is in its least active state is established from an advanced model of TSI variability 
\item This lower limit on grand minima TSI truncates the current range of the possible change in TSI since the Maunder minimum by more than half
\end{keypoints}

\begin{abstract}
How the solar electromagnetic energy entering the Earth’s atmosphere varied since pre-industrial times is an important consideration in the climate change debate. Detrimental to this debate, estimates of the change in total solar irradiance (TSI) since the Maunder minimum, an extended period of weak solar activity preceding the industrial revolution, differ markedly, ranging from a drop of 0.75 W m$^{-2}$ to a rise of 6.3 W m$^{-2}$. Consequently, the exact contribution by solar forcing to the rise in global temperatures over the past centuries remains inconclusive. Adopting a novel approach based on state-of-the-art solar imagery and numerical simulations, we establish the TSI level of the Sun when it is in its least active state to be 2.0$\pm$0.7 W m$^{-2}$ below the 2019 level. This means TSI could not have risen since the Maunder minimum by more than this amount, thus restricting the possible role of solar forcing in global warming.
\end{abstract}

\section*{Plain Language Summary}

How the amount of energy the Earth receives from the Sun varied since pre-industrial times is an important consideration in the climate change debate. Detrimental to this debate, it is not known whether the Sun grew brighter or dimmer since the 16th century, and by how much. As a consequence, the exact contribution by fluctuations in the brightness of the Sun to the rise in global temperatures over the past centuries remains controversial. It is established that the Sun was particularly inactive over much of the 16th century. Adopting a novel approach based on state-of-the-art solar imagery and computer models, we determined the brightness of the Sun when it is in its least active state possible. This places a strict limit on how much the Sun could have grown brighter since the lull in solar activity over the 16th century, restricting the possible role the Sun could have played in global warming.

\section{Introduction}

Climate change is one of the most pressing challenges facing humanity. The urgency is underscored by the recent pronouncement by the Intergovernmental Panel on Climate Change that greenhouse gas emissions will have to be halved by 2030 to keep global warming below the critical threshold of 1.5 °C \cite{ipcc18}. Numerical simulations of the Earth’s climate are central to the understanding of past climate and predictions of future trends. They require, as input, the time variation in the various agents of climate change, both anthropogenic and natural \cite{ipcc13}. A key natural driver is the variability in the electromagnetic energy reaching the Earth from the Sun \cite{gray10}. This is usually described in terms of solar irradiance, which is the solar radiative flux above the Earth’s atmosphere at the mean Earth-Sun distance. The long-term or secular trend in solar irradiance since pre-industrial times, and how this variability is distributed in wavelength are of particular interest for their relevance to the overall influence of the Sun on the Earth’s climate over the past centuries \cite{solanki13}. In this article, our focus will be on the particular issue that current estimates of the secular change in the wavelength-integrated total solar irradiance (TSI) since the 16th century diverge significantly, ranging from a drop of 0.75 W m$^{-2}$ to a rise of 6.3 W m$^{-2}$ \cite{tapping07,steinhilber09,schrijver11,shapiro11,judge12,coddington16,dasiespuig16,egorova18,wu18,lockwood20}.

Since the reliable measurement of solar irradiance only started in 1978 \cite{kopp14}, climate simulations rely on models of solar irradiance variability to provide the requisite historical solar forcing input. Except at the shortest and the longest timescales, where oscillations, convection and stellar evolution become relevant, solar activity is dominated by its magnetism, making it the prime candidate driver of solar irradiance variability at intermediate timescales \cite{solanki13}. Indeed, various studies have demonstrated the variation in solar irradiance at timescales of days to decades to be mainly the result of solar surface magnetism \cite{yeo14b,yeo17,shapiro17}. This is widely considered to also be the case at longer timescales of centuries and millennia \cite{solanki13}. The photospheric magnetic field is partially confined in kilogauss-strength magnetic concentrations, which manifest themselves as dark sunspots and bright faculae and network \cite{spruit83}. Solar irradiance fluctuates with the changing prevalence and distribution of these bright and dark magnetic structures on the Earth-facing solar hemisphere. Based on this knowledge, historical solar irradiance is reconstructed from proxies of solar magnetism by modelling the effect of photospheric magnetism on solar irradiance \cite{tapping07,steinhilber09,shapiro11,coddington16,dasiespuig16,egorova18,wu18}.

\begin{figure}
\noindent\includegraphics[width=\textwidth]{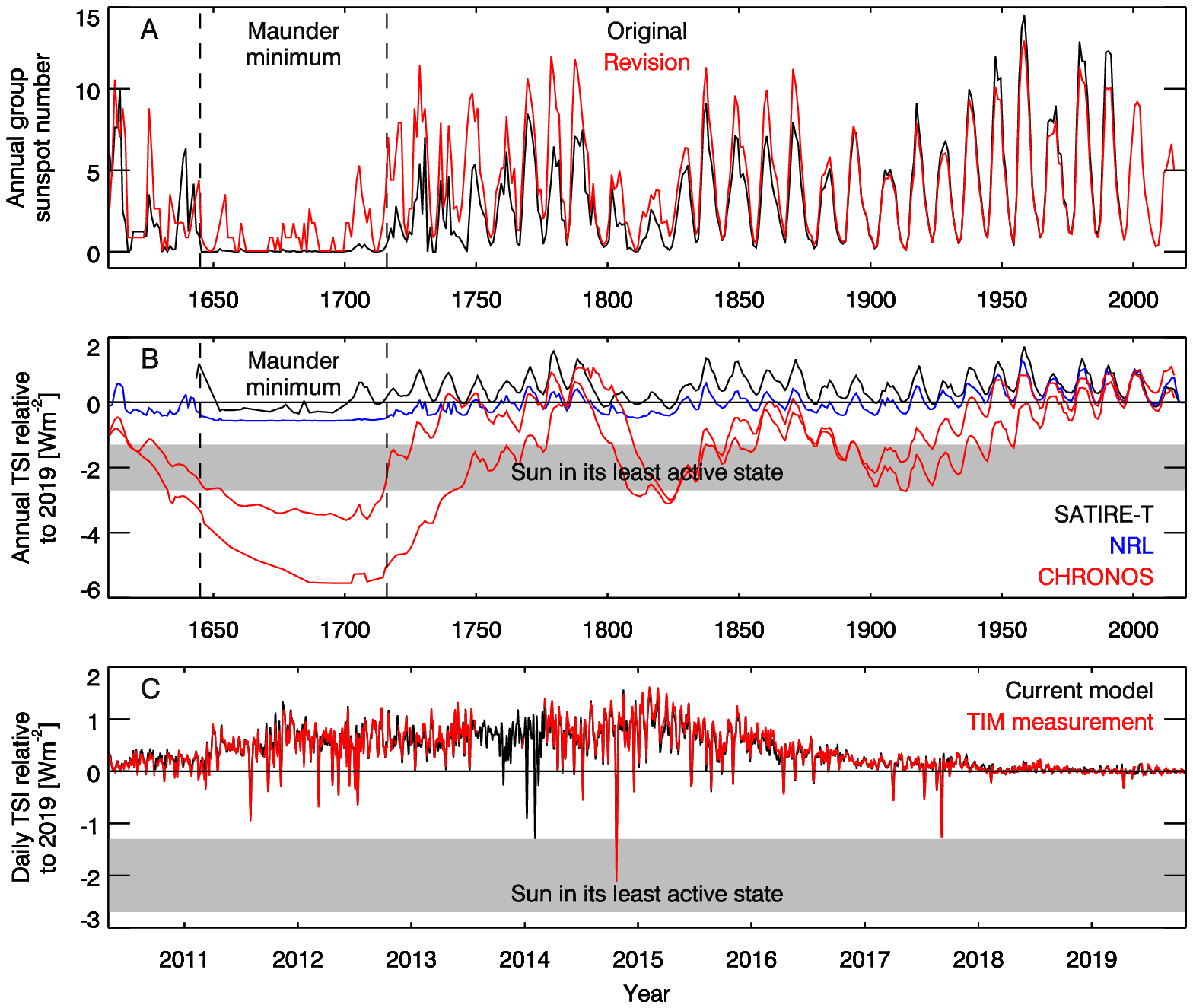}
\caption{(A) Solar activity since the Maunder minimum, the period within the vertical dashed lines, as indicated by the group sunspot number. The original time series by \citeA{hoyt98} (black) and the recent revision by \citeA{svalgaard16} (red) are depicted. (B) The reconstruction of TSI over the same period from the SATIRE-T \cite{wu18} (black), NRL \cite{coddington16} (blue) and CHRONOS models \cite{egorova18} (red), illustrating the uncertainty in the change in TSI since the Maunder minimum. (There are multiple CHRONOS TSI reconstructions, differing by the input data used. The time series indicating the weakest and strongest secular variability are depicted.) (C) The reconstruction of TSI since 2010 from the current study (black) and the measurements from SORCE/TIM \cite{kopp05} (red), which monitored TSI over the same period except between August 2013 and February 2014. The root-mean-square difference between the two daily time series is 0.067 W m$^{-2}$ and the Pearson’s correlation coefficient is 0.987, i.e., the model replicates about 97\% of the observed variability in the TIM record. We establish, with this model, that the TSI level of the Sun in its least active state is 2.0$\pm$0.7 W m$^{-2}$ below the 2019 level, indicated by the shaded range in B) and C). Since this is the lowest level TSI can take over solar grand minima, it implies TSI could not have risen since the Maunder minimum by more than this margin, excluding the acute secular variability indicated by CHRONOS.}
\label{tsifig}
\end{figure}

The various reconstructions of TSI extending back to pre-industrial times indicate divergent secular variability. The Maunder minimum is a solar grand minimum, a multi-decade period of weak solar activity, which occurred just before the industrial revolution (around 1645 to 1715, Fig. \ref{tsifig}A). According to the CHRONOS model \cite{egorova18}, between the Maunder minimum and the ongoing solar activity cycle minimum, TSI rose by 3.9 to 6.3 W m$^{-2}$ (Fig. \ref{tsifig}B). This is a recent revision of the model by \citeA{shapiro11}, which estimated a rise of 6$\pm$3 W m$^{-2}$. The revision was prompted by the assertion by \citeA{judge12} that \citeA{shapiro11} might have over-estimated this quantity by a factor of about two. The other model-based estimates \cite{tapping07,steinhilber09,coddington16,dasiespuig16,wu18} are markedly weaker and relatively consistent with one another, indicating a rise of between 0.3 and 0.9 W m$^{-2}$ (Fig. \ref{tsifig}B). In another recent study, \citeA{lockwood20} concluded, from an analysis of the measurements and models of TSI over the past two solar cycles, that it cannot be excluded from the uncertainty in this data that TSI might have been 0.4 W m$^{-2}$ lower to 0.75 W m$^{-2}$ higher during the Maunder minimum. Furthermore, there is a claim that the last cycle minimum in 2008 might be representative of the Maunder minimum \cite{schrijver11}. This implies that TSI rose since the Maunder minimum by the same margin as it did since 2008, about 0.1 W m$^{-2}$. By current estimates, since the Maunder minimum, TSI could have risen by more than 3 W m$^{-2}$ \cite{shapiro11,judge12,egorova18} or by less than 1 W m$^{-2}$ \cite{tapping07,steinhilber09,schrijver11,coddington16,dasiespuig16,wu18}, or even declined instead \cite{lockwood20}.

While there is clear evidence that greenhouse gas emissions play a significant role in global warming \cite{shangguan19}, quantifying its exact effects is less than straightforward due to, amongst other factors, the uncertainty in the secular variability in solar forcing. For certain climate models \cite{feulner11,fernandezdonato13,hind13,schurer14,luterbacher16}, assuming a 3 to 6 W m$^{-2}$ rise in TSI since the Maunder minimum produces a 0.3 to 0.4 °C greater increase in Northern Hemisphere temperatures over this period than if a rise in TSI of less than 1 W m$^{-2}$ were considered. A steep secular rise in TSI since the Maunder minimum points to solar forcing effecting a greater temperature rise over the past centuries and hence a smaller role by greenhouse gas emissions in driving climate change. It also alludes to a greater offset to global warming by the Sun if it were to enter a grand minimum state in the future. Unfortunately, due in part to the divergence between different climate models, it remains inconclusive from climate modelling if weak or strong solar forcing is more consistent with historical temperature reconstructions \cite{ljungqvist19}. There is a vital need to constrain TSI secular variability more tightly and to do so in a manner that circumvents the limitations of the earlier studies which led to the nagging uncertainty in this quantity.

To address this exigency, we make use of the recently-developed, novel model of solar irradiance variability reported by \citeA{yeo17} to constrain the rise in TSI since the Maunder minimum. The \citeA{yeo17} model is advanced in that it is the only one reported in the literature to not require any calibration to measured solar irradiance variability. Instead of estimating the change in TSI since the Maunder minimum by reconstructing TSI back to this time, the approach and aim here is to set an upper limit by establishing the TSI level corresponding to the Sun in its least active state. As this represents the lowest level TSI can take during grand solar minima, the difference to the present level represents the greatest possible rise in TSI since the Maunder minimum.

\section{The Least Active State of the Sun}
\label{leastactivestate}

Let us consider the question of what is the least active state of the Sun. Except at the shortest and longest timescales, solar activity is dominantly driven by its magnetism. The solar magnetic field is perpetuated by global and small-scale turbulent dynamo processes. The global dynamo refers to the physical processes that produce the cyclic regeneration of the large-scale solar magnetic field that underlies the solar cycle, including the manifestation of sunspots, faculae and network \cite{charbonneau20}. The small-scale turbulent dynamo or SSD denotes the interaction between solar convection and magnetic flux that yields the ubiquitous small-scale turbulent magnetic field on the solar surface, the so-termed internetwork magnetic field \cite{borrero17}. There is evidence to suggest that the internetwork magnetic field might also contribute to the magnetic network.

The dearth of sunspots during the Maunder minimum (Fig. \ref{tsifig}A) implies that the global dynamo is weak during this and other grand solar minima. At the same time, recent studies indicate that the internetwork magnetic field is unaffected by larger-scale magnetic structures \cite{lites11,rempel14} and does not vary with the solar cycle \cite{buehler13,lites14}, implying that the SSD is not or only weakly coupled to the global dynamo. We emphasize here that while existing studies found no clear evidence that the SSD might be coupled to the global dynamo, this possibility cannot be excluded. Under the assumption that the SSD is, as the cited studies suggest, not or only weakly coupled to the global dynamo, the diminishing strength of the global dynamo over grand solar minima has limited effect on the SSD. The Sun evolves on timescales of gigayears \cite{sackmann93} and the thermal relaxation timescale of the convection zone is in the order of $10^{5}$ years \cite{spruit82}. This means the convection zone, the engine of the SSD, is in essentially the same state today as during the Maunder minimum, or indeed over the past millennia. In summary, one, the global dynamo is weak during grand solar minima, two, the SSD is untethered to the global dynamo and therefore unaffected by any weakening of the latter, and three, solar convection, which drives the SSD, is invariant over the past millennia.

All considered, the scenario where the global dynamo is dormant, leaving just the SSD, represents the most inactive state the Sun can be in. In this state, there would be no sunspots or faculae due to the dormancy of the global dynamo, and the internetwork magnetic field, including the network it sustains, would extend over the entire solar surface. Assuming the SSD is, as the studies we have discussed here suggest, untethered to the global dynamo and invariant over the past millennia, the internetwork magnetic field would appear as it does today.

\section{The Brightness of the Sun in its Least Active State}
\label{analysis}

To establish the TSI level of the Sun in its least active state, and therefore the maximum possible secular rise since the Maunder minimum, we build on the recent, groundbreaking work by \citeA{yeo17} which presented the only reported model of solar irradiance variability to not require any calibration to measured variability. This advance, critical to our objective here, was achieved by deriving the intensity of certain solar surface features from three-dimensional (3D) magnetohydrodynamic (MHD) simulations of the solar surface and atmosphere. Such simulations have matured to such a level of physical realism that they are able to reproduce a wide range of observational constraints \cite{shelyag04,schussler08,danilovic10,danilovic13,danilovic16,afram11,rempel14,riethmuller14,yeo17,delpinoaleman18}.

Other models, for simplicity, either obtain the intensity of solar surface features from semi-empirical plane-parallel model solar atmospheres \cite{shapiro11,dasiespuig16,egorova18,wu18} or estimate their effect on solar irradiance by the linear regression of certain indices of solar magnetism to solar irradiance measurements \cite{tapping07,steinhilber09,coddington16}. As a direct consequence, they feature free parameters that have to be constrained by optimising the agreement between modelled and measured solar irradiance variability. This renders the model output, including the reconstructed secular variability, susceptible to the uncertainty in solar irradiance measurements. This is compounded by the fact that the linear regression of indices of solar magnetism to solar irradiance measurements is a crude simplification of the nuanced relationship between solar surface magnetism and irradiance variability \cite{yeo14c,yeo19}. Also, plane-parallel model atmospheres, the one-dimensional simplification of the spatially inhomogeneous solar atmosphere, cannot capture all the relevant physics \cite{uitenbroek11,holzreuter13}.

In an advance, the 3D MHD simulations-based model presented by \citeA{yeo17} reproduces measured TSI variability without any of these simplifications or calibration to the latter. The model does, however, make the assumption that the solar surface outside of sunspots, faculae and network (i.e., the internetwork) is field-free, neglecting the SSD and the internetwork magnetic field it maintains. The results of recent simulations \cite{rempel20}, indicating that the brightness of the internetwork changes with the strength of the internetwork magnetic field, suggest that this assumption might be problematic. In this study, we extend the \citeA{yeo17} model to include a physically realistic description of the internetwork by using an updated 3D MHD code to generate simulations of the SSD. This not only renders the model even more physical, but critically, it enables us to compute the TSI level of the Sun in its least active state, where the global dynamo is dormant, leaving just the SSD active (Sect. \ref{leastactivestate}). And the unique strength of the \citeA{yeo17} approach allows us to do so while circumventing the limitations of other models, which contributed to the gross uncertainty in TSI secular variability.

Based on the assumption that the variation in TSI at timescales greater than a day is dominantly-driven by solar surface magnetism, the \citeA{yeo17} model describes the variation in TSI due to the changing prevalence and distribution of bright and dark magnetic structures on the solar disc. The model has two inputs. The first is the solar disc coverage by magnetic features. These come in the form of faculae and network, classed together and termed collectively as faculae, and of sunspots. The disc coverage by faculae and sunspots is derived by identifying these solar surface features in full-disc longitudinal magnetograms and continuum intensity images from the HMI instrument onboard the SDO mission, launched in 2010 \cite{scherrer12}. The solar disc outside of sunspots and faculae (including network) is taken to be the internetwork. The second input is the emergent intensity from sunspots, faculae and the internetwork. Sunspot intensity is calculated from plane-parallel model atmospheres \cite{unruh99}, and facular and internetwork intensity from 3D model atmospheres using radiative transfer codes. Sunspots can be reasonably well-represented by plane-parallel model atmospheres such that their application to the model reproduces the effect of sunspots on TSI while maintaining its independence from calibration to measured solar irradiance variability. The 3D model atmospheres are based on MHD simulations of facular and internetwork regions executed with the MURaM code \cite{vogler07,rempel14,rempel20}. Importantly, while \citeA{yeo17} had represented the internetwork with field-free simulations, neglecting the SSD and the internetwork magnetic field, we make use of a more recent version of the MURaM code than was used in this earlier study to generate simulations that emulate SSD action and replicate the internetwork magnetic field.

\begin{figure}
\noindent\includegraphics[width=\textwidth]{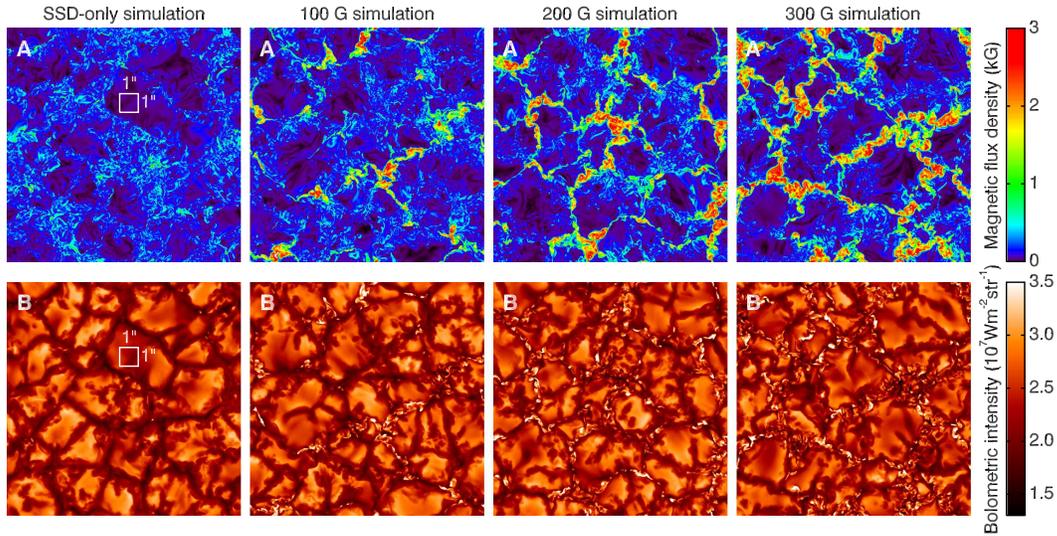}
\caption{3D MHD simulations of the solar surface and atmosphere, and the calculated emergent intensity. (A) Magnetic flux density at optical depth unity (i.e., at the solar surface) in a snapshot of each simulation run and (B) the corresponding bolometric image, calculated from each snapshot using a radiative transfer code. The simulation runs are set-up to emulate the small-scale turbulent dynamo, producing the ubiquitous small-scale turbulent magnetic field. Into three of the four runs, a uniform vertical magnetic field of 100 G, 200 G and 300 G, respectively was introduced. After these simulations reached statistical equilibrium, the imposed net magnetic flux aggregates in the intergranular lanes, forming the kilogauss-strength magnetic concentrations apparent as comparatively small bright features in the bolometric images. The white box in the left column denotes 1$\times$1 arcsec, corresponding to an area of approximately 725$\times$725 km on the Sun.}
\label{simfig}
\end{figure}

There are four simulation runs (Fig. \ref{simfig}A), setup to emulate SSD action \cite{vogler07,rempel14,rempel20}. The first run, representing the internetwork, is initiated with a weak, random magnetic field (root-mean-square field strength of 10$^{-3}$ G). This seed field is amplified and reorganised in the simulation into a small-scale turbulent field resembling the internetwork magnetic field. The other three runs, representing facular regions with varying levels of magnetic flux, were initiated with a uniform vertical magnetic field of 100 G, 200 G and 300 G, respectively. As in the first run, these runs emulate SSD action and replicate the internetwork magnetic field. However, having imposed net magnetic flux, these runs also form kilogauss-strength magnetic aggregations in the intergranular lanes, i.e., network and faculae-like features. Ten samples are taken of each run and the emergent intensity from each simulation snapshot, essentially a 3D model atmosphere, is computed (Fig. \ref{simfig}B). Internetwork intensity is given by the mean intensity of the SSD-only run snapshots, and facular intensity by the intensity of the facular features apparent in the snapshots of the facular runs. MURaM simulations are well-suited for the current purpose, having been demonstrated in multiple studies to reproduce the observed properties of the internetwork magnetic field \cite{schussler08,rempel14,danilovic10,danilovic16,delpinoaleman18} and of faculae and network \cite{shelyag04,afram11,danilovic13,riethmuller14,yeo17}.

Given the surface coverage by sunspots and faculae at a particular time, the model computes the corresponding TSI level by assigning to each point on the solar disc the appropriate intensity depending on whether it is in a sunspot, in faculae or the internetwork. Using an HMI longitudinal magnetogram and the concurrent continuum intensity image from each day between 30 April 2010, when the instrument started regular operation, and 31 December 2019 as input into our TSI model, we reconstruct the daily TSI over the intervening period (Fig. \ref{tsifig}C). The reconstruction closely reproduces measured TSI variability. The Sun is at its least active when the global dynamo is dormant and the internetwork magnetic field maintained by the SSD extends over the entire solar surface (Sect. \ref{leastactivestate}). And since the SSD is likely untethered to the global dynamo and invariant over the past millennia, the TSI level of the Sun in this state emerges from the model by taking the entire solar disc to resemble the present-day internetwork. The absolute scale of the model is set to that of the SORCE/TIM TSI record \cite{kopp05}, the radiometry of which is widely regarded to be the most reliable \cite{fehlmann12}. To this end, we normalise the reconstruction of daily TSI over the last decade to the TIM record and apply the same normalisation factor to the TSI level of the Sun in its least active state, arriving at the final value of 1358.7$\pm$0.7 W m$^{-2}$ (shaded range, Figs. 1B and 1C). We emphasize that this normalisation merely sets the absolute scale of the model to that of the TIM record, the TSI level of the Sun in its least active state relative to the present level is unaffected. The associated error, 0.7 W m$^{-2}$, denotes the uncertainty in this quantity from the uncertainty in the simulation of the internetwork magnetic field. This is a conservative estimate derived by repeating the above analysis for the extreme scenario that the internetwork is field-free.

Since the TSI level of the least active Sun, 1358.7$\pm$0.7 W m$^{-2}$ is the lowest level TSI can take during grand solar minima, the difference to the 2019 level, 2.0$\pm$0.7 W m$^{-2}$ (shaded range, Figs. 1B and 1C), represents the maximum possible rise in TSI since the Maunder minimum. The TSI model and the above analysis are further detailed in the Supporting Information.

The occurrence of numerous and/or sizable sunspots can produce large dips in TSI, even below the level of the least active Sun, 1358.7$\pm$0.7 W m$^{-2}$, such as in October 2014 (Fig. \ref{tsifig}C). It is unlikely, however, that sunspots might have suppressed the secular trend in TSI over the Maunder minimum below this level. Firstly, there was a dearth of sunspots during the Maunder minimum (Fig. \ref{tsifig}A). Secondly, darkening by a sunspot is a short-term phenomenon and not a state the Sun can sustain over the duration of a solar cycle, much less a grand minimum.

\section{Comparison to Other Estimates of the Change in TSI Since the Maunder Minimum}

The cap on the rise in TSI since the Maunder minimum derived here, 2.0$\pm$0.7 W m$^{-2}$, encloses all the current estimates \cite{tapping07,steinhilber09,schrijver11,shapiro11,judge12,coddington16,dasiespuig16,egorova18,wu18,lockwood20} except the range of 3 to 6.3 W m$^{-2}$ indicated by \citeA{shapiro11}, the reassessment of their model by \citeA{judge12}, and CHRONOS \cite{egorova18} (Fig. \ref{tsifig}B). We consider the upper limit established here to be more reliable than the existing estimates. As argued in Sect. \ref{analysis}, the \citeA{yeo17} modelling approach circumvents the limitations of the simpler models employed in the earlier works. And as we will argue next, the model atmospheres employed by \citeA{shapiro11} and CHRONOS are ill-suited for estimating TSI secular variability via their methodology. In these two models, the intensity of the quiet Sun is assumed to vary with solar activity between two defined limits, and this forms the main contribution to the reconstructed secular variability. (The quiet Sun refers to the solar surface outside of sunspots and faculae, i.e., network and internetwork.) In \citeA{shapiro11}, the lower and upper limits are given by the FAL-A and FAL-C plane-parallel model solar atmospheres \cite{fontenla99}, respectively. The FAL-A and FAL-C models correspond to quiet Sun regions with different amounts of magnetic flux and therefore different intensities. \citeA{judge12} raised questions over the use of the FAL-A model as the lower limit, leading CHRONOS to employ an amalgamation of the FAL-A and FAL-C models, termed PW-B, as the lower limit instead \cite{egorova18}.

The FAL model atmospheres are, however, not designed for their intended purpose in \citeA{shapiro11} and CHRONOS. As noted earlier, plane-parallel representations of the spatially inhomogeneous solar atmosphere cannot capture all the relevant physics. Multiple studies demonstrated that plane-parallel model atmospheres cannot fully return the true average intensity spectra of the regions they represent \cite{uitenbroek11,holzreuter13}. More critically, while solar electromagnetic radiation and consequently TSI variability is produced mainly in the photosphere, the photospheric layers of the FAL model atmospheres, and therefore also those of PW-B, are poorly constrained \cite{vernazza81}. This is due to the fact that the FAL model atmospheres are based on ultraviolet measurements formed largely in the chromosphere. As an example, if the PW-B model were to be made a mere 10 K warmer, a margin smaller than the accuracy of this and other such model atmospheres, CHRONOS TSI secular variability would decrease by a factor of about five (see Supporting Information). Consequently, the FAL models are not accurate enough to support the acute TSI secular variability indicated by \citeA{shapiro11} and CHRONOS. The approach taken here, by relying on 3D model atmospheres based on realistic simulations of the solar surface and atmosphere that reproduces a wide range of photospheric observational constraints \cite{shelyag04,schussler08,danilovic10,danilovic13,danilovic16,afram11,rempel14,riethmuller14,yeo17,delpinoaleman18}, circumvents these limitations.

\section{Conclusion}

There is significant uncertainty in the historical secular variability in solar irradiance, with current estimates of the change in TSI since the Maunder minimum ranging from a drop of 0.75 W m$^{-2}$ to a rise of 6.3 W m$^{-2}$. This in turn introduces considerable uncertainty in the contribution of the Sun to climate change. By demonstrating that the Sun in its least active state is at the most 2.0$\pm$0.7 W m$^{-2}$ dimmer than it was over 2019, we greatly constrain the change in TSI since the Maunder minimum by ruling out rises above this margin. This constraint on TSI secular variability over the past centuries, by limiting the possible contribution by solar forcing to global temperature changes, will be a boon to the climate change assessment.

\begin{acronyms}
\acro{3D}       3-Dimensional
\acro{CHRONOS}  Code for the High spectral ResolutiOn recoNstructiOn of Solar irradiance
\acro{FAL-A}    Fontenla, Avrett and Loeser model A
\acro{FAL-C}    Fontenla, Avrett and Loeser model C
\acro{HMI}      Helioseismic and Magnetic Imager
\acro{IPCC}     Intergovernmental Panel on Climate Change
\acro{MHD}      MagnetoHydroDynamic
\acro{MURaM}    Max Planck Institute for Solar System Research and University of Chicago Radiation MHD code
\acro{NRL}      Naval Research Laboratory
\acro{PW-B}     Present Work model B
\acro{SATIRE-T} Spectral And Total Irradiance REconstruction for the Telescope era
\acro{SDO}      Solar Dynamics Observatory
\acro{SORCE}    SOlar Radiation and Climate Experiment
\acro{SSD}      Small-Scale Dynamo
\acro{TIM}      Total Irradiance Monitor
\acro{TSI}      Total Solar Irradiance
\end{acronyms}

\acknowledgments

We are grateful to the anonymous referee and M. Lockwood of the University of Reading for their constructive inputs. We also thank M. van Noort of the Max Planck Institute for Solar System Research and the \#335 science team supported by the International Space Science Institute for the useful discussions. K.L.Y., S.K.S. and N.A.K. received funding from the German Federal Ministry of Education and Research (Project No. 01LG1909C). This work received funding from the European Research Council under the European Union Horizon 2020 research and innovation program (K.L.Y. and S.K.S., Grant Agreement No. 695075; L.S.A., A.I.S. and V.W., Grant Agreement No. 715947). S.K.S. is supported by the BK21 plus program of the National Research Foundation funded by the Ministry of Education of Korea. M.R. is supported by the National Aeronautics and Space Administration (LWS grant NNX16AB82G). R.V.T. is supported by the Science and Technology Facilities Council (consolidated grant ST/S000372/1). This material is based upon work supported by the National Center for Atmospheric Research (NCAR), which is a major facility sponsored by the National Science Foundation under Cooperative Agreement No. 1852977. We acknowledge high-performance computing support from Cheyenne (doi: 10.5065/D6RX99HX) provided by NCAR’s Computational and Information Systems Laboratory. We declare that none of the authors have any competing interests. The group sunspot number record is available at www.sidc.be/silso/groupnumberv3 and the SORCE/TIM TSI record at lasp.colorado.edu/lisird/. The TSI model makes use of solar imagery from the NASA SDO mission, available at jsoc.stanford.edu. The model output TSI reconstruction is available at www2.mps.mpg.de/projects/sun-climate/data.html.

\nocite{anders89,bellotrubio02,couvidat12,frutiger00,hotta15,kurucz92,mcclintock05,norris17,piskunov95,rogers96,solanki98,tagirov17,topka92,unruh99,vogler04,vogler07,yeo14a,yeo14b,yeo13}

\bibliography{references}

\end{document}